\documentclass[aps,prl,twocolumn,showpacs,superscriptaddress]{revtex4-1}
\usepackage[colorlinks ,linkcolor=blue,anchorcolor=blue,citecolor=blue,urlcolor=blue]{hyperref}
\usepackage{graphicx}
\usepackage{epstopdf}
\usepackage{soul, color, xcolor}

\def\be{\begin{equation}} \def\ee{\end{equation}}
\def\bea{\begin{eqnarray}} \def\eea{\end{eqnarray}}

\begin{document}

\title{Multiple types of spin textures and robust valley physics in MP$_2$X$_6$}
\author{Li Liang}
\thanks{Current address: School of Physics, Southeast University, Nanjing 211189, China} 
\affiliation{School of Physics and Technology, Nanjing Normal University, Nanjing 210023, China}
\affiliation{Center for Quantum Transport and Thermal Energy Science,
Nanjing Normal University, Nanjing 210023, China} 
\author{Zhichao Zhou}
\email{zczhou@njnu.edu.cn}
\affiliation{School of Physics and Technology, Nanjing Normal University, Nanjing 210023, China}
\author{Jie Zhang}
\affiliation{School of Physics and Technology, Nanjing Normal University, Nanjing 210023, China}
\affiliation{Center for Quantum Transport and Thermal Energy Science,
Nanjing Normal University, Nanjing 210023, China} 
\author{Xiao Li}
\email{lixiao@njnu.edu.cn}
\affiliation{School of Physics and Technology, Nanjing Normal University, Nanjing 210023, China}
\affiliation{Center for Quantum Transport and Thermal Energy Science,
Nanjing Normal University, Nanjing 210023, China}

\begin{abstract} 
Both spin textures and multiple valleys in the momentum space have attracted great attentions due to their versatile applications in spintronics and valleytronics. 
It is highly desirable to realize multiple types of spin textures in a single material and further couple the spin textures to valley degree of freedom.  
Here, we study electronic properties of SnP$_{2}$Se$_{6}$ monolayer by first-principles calculations. 
The monolayer exhibits rare Weyl-type and Ising-type spin textures at different valleys, which can be conveniently expressed by electron and hole dopings, respectively. 
Besides valley-contrasting spin textures, Berry-curvature-driven anomalous Hall currents and optical selectivity are found to be valley dependent as well. 
These valley-related properties also have generalizations to SnP$_{2}$Se$_{6}$ few-layers and other MP$_{2}$X$_{6}$.   
Our findings open new avenue for exploring appealing interplay between spin textures and multiple valleys, and designing advanced device paradigms based on spin and valley degrees of freedom. 
\end{abstract} 
\pacs{} 
\maketitle

{\color{blue}\textit{Introduction.}} -- The spin-orbit coupling gives rise to spin splittings and spin-momentum locking in electronic structures of noncentrosymmetric crystal materials \cite{Manchon2015, Chen2021}. 
The spin textures in the momentum space enable a large number of spin-related phenomena, which have enormous applications in spintronics. 
Two extensively studied types of spin-orbit couplings are the Rashba-type and Dresselhaus-type ones. 
They have been found in a wide range of materials, such as heterojunctions and surface systems with structure inversion asymmetry \cite{Vasko1979, Bychkov1984, Nitta1997}, and three-dimensional materials with bulk inversion asymmetry \cite{Dresselhaus1955}.  
In contrast, Ising-type and Weyl-type spin-orbit couplings were proposed in recent years and they have fewer studies and material realizations \cite{Zhu2011, Xiao2012, Hirayama2015, Sakano2020, Gatti2020, Krieger2024}. 
However, Ising-type and Weyl-type spin textures are likely to result in distinct physical consequences that can't be realized in previous Rashba-type and Dresselhaus-type spin textures, e.g. new types of superconductivity \cite{Yuan2014, Lu2015, Wang2019, Falson2020, Liu2020, Xie2020, ZhangE2023, Fischer2023}  and spin-orbit torques  \cite{Shao2016, Li2020, Kang2021, He2021, Burkov2023}. 
Therefore, material candidates exhibiting Ising-type or Weyl-type spin textures are highly sought after. 
Taking one step further, a single material with both types of spin textures is unprecedented and worth being discovered.

On the other hand, valleys, energy extrema in electronic band structures, have been emerged as a new degree of freedom \cite{Xiao2007, Schaibley2016, Vitale2018, Liu2019}. 
In two-dimensional multi-valley materials, inequivalent valleys exhibit contrasting physical properties, such as opposite anomalous Hall currents and optical circular dichroism \cite{Xiao2007, Yao2008}.  
 Moreover, considering spin degrees of freedom, the spin-valley coupling also appears in electronic structures, i.e. different valleys exhibit opposite Ising-type spin splittings  \cite{Xiao2012, LiS2020}. 
These valley-contrasting properties provide potential uses in advanced information technology. 
Given that both spin textures and valley degree of freedom are of importance in physical phenomena and applications, an intriguing question arises as to whether different types of spin textures can be coupled to inequivalent valleys. 
If so, corresponding electron systems will directly inherit merits from both spin textures and valley degree of freedom. 
Furthermore, the spin-related effects are expected to be tuned through the valley degree of freedom and vice versa.

In this work, taking the SnP$_2$Se$_6$ monolayer as a typical example, we study electronic energy valleys, spin textures, and their couplings in van der Waals layered MP$_2$X$_6$ (M=Sn, Ge; X=S, Se). 
 Recently, SnP$_{2}$Se$_{6}$ have been synthesized in experiments, highlighting its outstanding features for electronic and optoelectronic applications \cite{Zhu2023, Sangwan2024, Zhu2024}. 
Our first-principles calculations reveal that the SnP$_{2}$Se$_{6}$ monolayer behaves as an indirect-band-gap semiconductor, with valence band maxima at $K_{\pm}$ and conduction band minimum at $\Gamma$. 
Interestingly, the band edge at $K_{\pm}$ and $\Gamma$ display Ising-type and Weyl-type spin textures, respectively. 
The distinct spin textures at different valleys can be selectively manipulated through e.g. the carrier doping. 
Moreover, we identify valley-contrasting Berry curvature and optical circular dichroism at $K_{\pm}$ valleys, which,  together with the spin textures, remain robust across varied layer numbers. 
The coexistence of various spin textures and robust valley physics in a single material provides a powerful platform for studying spin-valley-related physics, and presents promising avenues for the development of high-performance valleytronic and spintronic devices.

\begin{figure}[htb]
\includegraphics[width=7 cm]{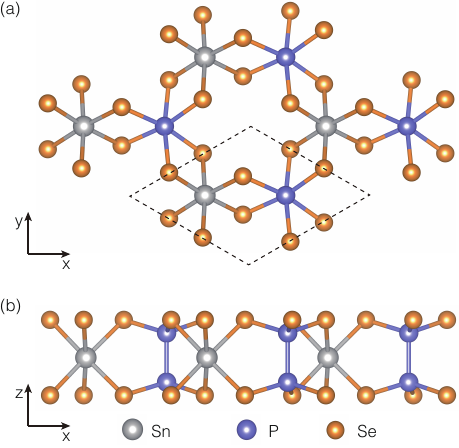}
\caption{ Atomic structure of the SnP$_{2}$Se$_{6}$ monolayer. 
(a) The top view and (b) the side view. 
Grey, blue, and orange balls stand for Sn, P and Se atoms, respectively.  
The unit cell is bounded by dashed lines, exhibiting a hexagonal crystal structure. 
}
\label{fig01}
\end{figure}

{\color{blue}\textit{Crystal structures.}} -- By the first-principles calculations, we investigate the crystal structure and electronic properties of the SnP$_{2}$Se$_{6}$ monolayer, the calculation details is provided in Supplemental Material (SM) \cite{SMsnp2se6}. 
Figs. \ref{fig01} (a) and (b) show its relaxed crystal structure from both the top and side views, respectively. 
In a unit cell, SnP$_2$Se$_6$ consists of one Sn$^{4+}$ ion and one [P$_{2}$Se$_{6}$]$^{4-}$ anion cluster. 
The anion cluster is isostructural to the molecular ethane in a staggered conformation. 
The monolayer has a point group of $D_{3}$, which includes three-fold rotational symmetry but excludes spatial inversion symmetry and horizontal mirror symmetry.  
 The mirror symmetry breaking in the monolayer arises from the staggered conformation of the anion cluster. 
These symmetry breakings allow for the appearance of spin splittings and valley-contrasting properties in this monolayer, which will be discussed later.

Moreover, the in-plane lattice constant, $a$, of the SnP$_{2}$Se$_{6}$ monolayer is computed to be 6.50 \AA, which is consistent with previous theoretical and experimental results \cite{LinM2019, Zhu2023}. 
The bond lengths of the P-P atoms and P-Se atoms are 2.26 \AA\space and 3.92 \AA, respectively. 
The thickness of the monolayer, i.e., the distance between top and bottom Se atomic layers,  is 3.54 \AA.

\begin{figure}[bht]
\includegraphics[width=8.7 cm]{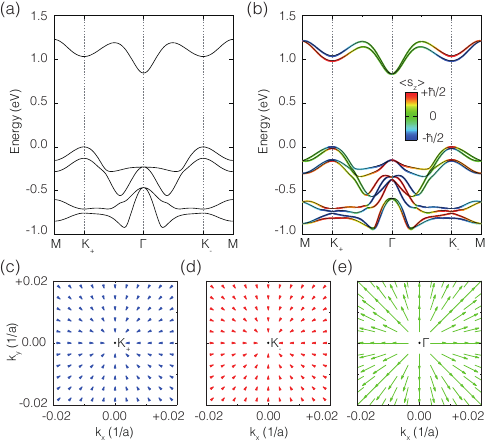}
\caption{ Electronic band structures and spin textures of the SnP$_2$Se$_6$ monolayer. 
(a) Nonrelativistic electronic band structure. 
 (b) Electronic band structure with considering the spin-orbit coupling. 
(c)-(e) Spin textures in small regions of the momentum space centered at $K_{+}$, $K_{-}$ and $\Gamma$ valleys, respectively. 
The magnitudes of out-of-plane spin components of electronic states are coded by color in (b)-(e), 
while the lengths of the arrows in (c)-(e) are proportional to the magnitudes of in-plane spin components. 
}
\label{fig02}
\end{figure}

{\color{blue}\textit{Distinct spin textures at valleys.}} -- We then study electronic band structures of the SnP$_2$Se$_6$ monolayer. 
 Fig. \ref{fig02} (a) and (b) demonstrate its band structures using the Perdew-Burke-Ernzerhof (PBE) functional \cite{Perdew1996}, without and with the spin-orbit coupling, respectively. 
 As illustrated in Fig. \ref{fig02} (a), the band structure without the spin-orbit coupling exhibits the conduction band minimum at $\Gamma$ and valence band maxima at $K_{\pm}$. 
 The monolayer is thus an indirect-band-gap semiconductor, with a calculated global gap of 0.84 eV.   
  Besides, it is found that there are local minima of the conduction band at $K_{\pm}$. 
 Considering the local conduction band minima and global valence band maxima at $K_{\pm}$, 
 the direct band gap at $K_{\pm}$ is computed to be 1.03 eV. 
 Moreover, the extrema of electronic states at $K_{+}$ and $K_{-}$ are equal, which is protected by the time-reversal symmetry \cite{Xiao2012}. 
Therefore, besides the $\Gamma$ valley in the conduction band, $K_{\pm}$ points correspond to two degenerate but inequivalent valleys for both valence and conduction bands.

Taking into account the spin-orbit coupling, the band structure in Fig. \ref{fig02} (b) exhibits spin degeneracy splittings, while the valley degeneracy between $K_{\pm}$ is well kept. 
We first focus on the spin splittings at $K_{\pm}$ valleys. 
According to spin textures in Figs. \ref{fig02} (c) and (d), it is found that the out-of-plane spin component is dominant at $K_{\pm}$ valleys for both valence and conduction bands, while the in-plane spin components are negligible.  
When moving away from $K_{\pm}$ valleys, the in-plane components are increased. 
At $K_{+}$ valley, the upper one of the spin-splitted valence (conduction) band states is spin-down, while the lower one is spin-up. 
As for $K_{-}$ valley, the upper and lower states of the valence (conduction) band are spin-up and spin-down, respectively.   
Further considering the valley degeneracy, the spin splittings at $K_{\pm}$ valleys thus have the same magnitude but opposite signs, indicating the spin-valley coupling protected by the time-reversal symmetry.    
The magnitudes of spin splittings at $K_{\pm}$ valleys are computed to be 16.0 meV for the valence band and 59.8 meV for the conduction band. 
Moreover, there is no crossing points between spin-splitted bands in the neighborhood of $K_{\pm}$ valleys. 
The band dispersion and dominating out-of-plane spin components indicate that the spin textures is Ising-type at $K_{\pm}$ valleys, similar to the case of two-dimensional transition metal dichalcogenides \cite{Xiao2012}. 
This kind of spin texture can be well described by a Ising-type spin-orbit coupling Hamiltonian, $H_{\text{soc}}\propto\tau s_{z}$, where $\tau=\pm 1$ corresponds to $K_{\pm}$ valleys and $s_{z}$ is the  Pauli matrix for the out-of-plane spin. 
The Hamiltonian gives rise to valley-dependent and momentum-independent spin splittings, similar to our first-principles calculation results.

In stark contrast to $K_{\pm}$ valleys, the conduction band edge at $\Gamma$ valley exhibits distinct band dispersion and spin texture. 
It is found that there is a lateral shift between the spin-splitted conduction bands in the momentum space, and these two bands cross with each other at $\Gamma$. 
The spin textures in Fig. \ref{fig02} (e) demonstrate that spins of the conduction band states are in-plane around $\Gamma$ valley. 
For the lower electronic states of the spin-splitted conduction bands, the in-plane spin directions are parallel to the crystal momentum, while the spins of the upper electronic states are antiparallel to the crystal momentum, which indicates a radial spin texture. 
This kind of spin texture at $\Gamma$ valley is reminiscent of the Hamiltonian of two-dimensional Weyl fermions, $H_{\text{soc}}\propto k_{x} s_{x} + k_{y} s_{y} $, where $s_{x,y}$ are Pauli matrices for real in-plane spins and $k_{x,y}$ are in-plane crystal wave vectors.  
The Weyl Hamiltonian produces a momentum-dependent radial spin texture like our first-principles calculations.

The above Ising-type spin texture at $K_{\pm}$ valleys and Weyl-type spin texture at $\Gamma$ valley can be understood from the view of symmetry. 
The absence of the inversion symmetry enables the appearance of spin splittings in nonmagnetic materials. 
Different from transition metal dichalcogenides with the horizontal mirror \cite{Xiao2012}, the mirror symmetry breaking in the SnP$_2$Se$_6$ monolayer makes the in-plane spin components possible. 
Moreover, the little groups at $\Gamma$ and $K_{\pm}$ are $D_{3}$ and $C_{3}$, respectively. 
The $D_{3}$ symmetry excludes the Rashba-type spin texture and allows for the Weyl-type spin texture at $\Gamma$ valley \cite{Acosta2021}. 
In contrast, the $C_{3}$ symmetry ensures that the spins have only out-of-plane component at $K_{\pm}$ \cite{Xiao2012}.

The simultaneous presence of different kinds of spin textures in the same material is intriguing.
Firstly, given that spin textures are associated with rich spin-related phenomena, different spin splittings are expected to produce distinct outcomes in the same material.  
Secondly, it is convenient for the SnP$_2$Se$_6$ monolayer to selectively express different kinds of spin textures. 
Since the Ising-type and Weyl-type spin textures appear at distinct valleys of the valence and conduction bands, they can be selected by hole doping and electron doping, respectively.  
More methods for valley polarization are also likely to be adopted to exploit different types of spin textures. 
For example, the spin-related photocurrents may be generated at valleys by optical fields with different polarizations \cite{Lu2016}.  
Thirdly, material examples with Ising- and Weyl-type spin textures are much rarer than those with Rashba- and Dresselhaus-type spin textures. 
The Ising-type spin texture was recently discovered in valley materials \cite{Xiao2012, LiS2020}, and the Weyl-type spin texture was merely found in trigonal group-VI elements, tellurium and selenium \cite{Hirayama2015, Sakano2020, Gatti2020}, and the chiral topological semimetal, PtGa \cite{Krieger2024}. 
The materials with both types of spin textures are conspicuously absent. 
Considering that SnP$_2$Se$_6$ has been realized experimentally \cite{Zhu2023, Sangwan2024, Zhu2024}, the monolayer with both spin textures at different valleys provides a fertile playground for studying spin and valley physics.

In addition, with the spin-orbit coupling, the size of the global band gap becomes 0.83 eV, and those of the direct band gaps at $K_{\pm}$ valleys become 0.98 eV. 
To support the first-principles results from the PBE functional, we also calculated the band structure of the SnP$_2$Se$_6$ monolayer using more accurate Heyd-Scuseria-Ernzerhof (HSE) screened hybrid functional method \cite{Heyd2003}. 
The HSE method shows similar band structure and spin textures with PBE results, except some quantitative differences, which are provided in SM \cite{SMsnp2se6}.

\begin{figure}[htb]
\includegraphics[width=8.5 cm]{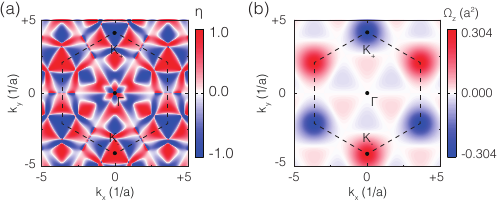}
\caption{ Valley-contrasting optoelectronic properties in the SnP$_2$Se$_6$ monolayer.
(a) The momentum-resolved degree of circular polarization. 
(b) The momentum-resolved Berry curvature.  
The first Brillouin zone is bounded by dashed lines. 
}
\label{fig03}
\end{figure}

{\color{blue}\textit{Optoelectronic properties at $K_{\pm}$ valleys.}} -- Given that there are degenerate but inequivalent valleys at $K_{\pm}$, we investigate potential contrasting optoelectronic properties at $K_{\pm}$ valleys in the SnP$_2$Se$_6$ monolayer. 
To explore the possibility of selectively exciting $K_{\pm}$ valleys and realizing dynamical valley polarization, the circular dichroism of the monolayer is studied. 
Fig. \ref{fig03} (a) illustrates the momentum-resolved degrees of circular polarization, $\eta(\textbf{\emph{k}})$, between the upper bands of both spin-splitted valence and conduction bands that have the same spin direction at $K_{\pm}$. 
$\eta(\textbf{\emph{k}})$ gives relative absorption rates of left- and right-circularly polarized light at a certain $\textbf{\emph{k}}$. 
It is seen that $\eta(\textbf{\emph{k}})$ has nonvanishing values in the neighborhood of $K_{\pm}$ valleys, but with opposite signs. 
At $K_{\pm}$, $\eta$ is computed to be $\pm 1$, indicating that the left- and right-circularly polarized lights are exclusively absorbed, respectively. 
Therefore, the SnP$_{2}$Se$_{6}$ monolayer exhibits a perfect valley-selective circular dichroism.

On the other hand, the momentum-resolved Berry curvature, $\Omega_{z}(\textbf{\emph{k}})$, is also calculated, which is illustrated in Fig. \ref{fig03}(b). 
Similar to the circular dichroism, the Berry curvature has considerable, opposite values around $K_{\pm}$ valleys. 
The valley-contrasting Berry curvature drives opposite anomalous Hall currents with opposite spins, under the actions of in-plane electric fields and hole doping. 
That is, both valley Hall effect and spin Hall effect can emerge in this monolayer. 
Considering the optical selectivity mentioned above, Berry-curvature-driven anomalous Hall current can also be generated by optical means.  
Therefore, besides the valley-selective spin textures, different valleys in the SnP$_{2}$Se$_{6}$ monolayer exhibit contrasting optoelectronic properties.

\begin{figure}[htb]
\includegraphics[width=8.6 cm]{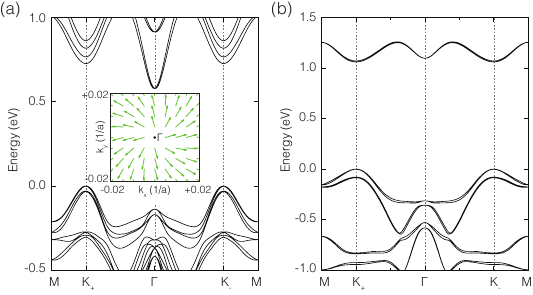}
\caption{ Electronic properties of the SnP$_2$Se$_6$ bilayer and GeP$_2$S$_6$ monolayer. 
(a) Electronic band structure of the SnP$_2$Se$_6$ bilayer with the spin-orbit coupling, 
of which the inset shows the spin texture in the neighborhood of $\Gamma$.  
(b) Electronic band structure of the GeP$_2$S$_6$ monolayer with the spin-orbit coupling.
}
\label{fig04}
\end{figure}

{\color{blue}\textit{Discussions}} --  Besides the monolayer form, van der Waals layered SnP$_{2}$Se$_{6}$ bulk and fewlayers have also been synthesized  \cite{Zhu2023, Sangwan2024, Zhu2024}. 
The bulk SnP$_{2}$Se$_{6}$ has a 3R interlayer stacking order, where the inversion symmetry is absent as well. 
The few-layer SnP$_{2}$Se$_{6}$ has also been found to be non-centrosymmetric by giant second-harmonic-generation activity \cite{Zhu2023, Sangwan2024}. 
While valley-contrasting optoelectronic properties in well-known valley materials, e.g. transition-metal dichalcogenides, exhibit odd-even dependencies \cite{Zhu2023}, the robustness of valley physics in SnP$_{2}$Se$_{6}$ is worth further exploring.   
 We then calculated valley-related properties of the SnP$_{2}$Se$_{6}$ bilayer. 
The most stable bilayer configuration is found to be a AB-stacked one (see SM \cite{SMsnp2se6}), which is similar to the interlayer stacking in the bulk form. 
Fig. \ref{fig04} (a) shows the electronic band structure of the bilayer, where the valence band maxima and conduction band minimum are located at $K_{\pm}$ valleys and $\Gamma$ valley, respectively, like the monolayer form. 
The spin textures at $K_{\pm}$ valleys for the valence band are Ising-type ones with opposite spin directions. 
As for the $\Gamma$ valley of the conduction band, there is a hybridization of Weyl-type and Rashba-type spin textures, as shown in the inset of Fig. \ref{fig04} (a), which is distinct from the case of the monolayer. 
The hybridization results from the symmetry reduction from $D_{3}$ to $C_{3}$ at $\Gamma$ \cite{Acosta2021}. 
The spin textures in the bilayer thus keep valley dependences, and even provide richer patterns. 
Moreover, according to the calculations of optical selectivity and Berry curvature, the bilayer also exhibits circular dichroism and valley-contrasting anomalous Hall transports. 
The above valley properties also apply to the SnP$_{2}$Se$_{6}$ trilayer \cite{SMsnp2se6}.  
Therefore, the valley physics in SnP$_{2}$Se$_{6}$ is more robust, compared with well-known valley materials.

Moreover, since SnP$_{2}$Se$_{6}$ belongs to the MP$_{2}$X$_{6}$ family, the intriguing characteristics found in the SnP$_{2}$Se$_{6}$ monolayer is expected to have more material generalizations.  
We further study electronic properties of a GeP$_{2}$S$_{6}$ monolayer. 
Similar to SnP$_{2}$Se$_{6}$, the GeP$_{2}$S$_{6}$ monolayer also exhibits valley-contrasting spin texture and optoelectronic properties (see SM \cite{SMsnp2se6}). 
On the other hand, the monolayer is computed to be a direct-band-gap semiconductor, with the valence band maxima and conduction band minima being both located at $K_{\pm}$ in Fig. \ref{fig04} (b), which is different from the SnP$_{2}$Se$_{6}$ monolayer. 
That is, as for the conduction band, the hole doping can be used to express the $\Gamma$ valley of the SnP$_{2}$Se$_{6}$ and $K_{\pm}$ valleys of the GeP$_{2}$S$_{6}$, respectively. 
The MP$_{2}$X$_{6}$ family thus provide more room to modulate the relative energies of these valleys and exploit the valley degree of freedom.
Besides, more external stimuli, e.g. applied magnetic, electric, and strain fields, are also likely to play the similar role in tuning the valley structure and associated properties \cite{Schaibley2016, Vitale2018, Liu2019, Li2020}.

{\color{blue}\textit{Summary}} -- To conclude, electronic properties of the SnP$_{2}$Se$_{6}$ monolayer are investigated by first-principles calculations. 
A Weyl-type spin texture and opposite Ising-type spin textures are discovered at different valleys. 
These types of spin textures are very rare, let alone realized in the same material. 
They, associated with plentiful spin-related phenomena, can be selectively expressed by carrier doping.   
 With the help of the spin-resolved angle-resolved photoemission spectroscopy \cite{Sakano2020, Gatti2020, Krieger2024}, the spin textures are also expected to be directly detected in experiments.
Besides, different valleys in the SnP$_2$Se$_6$ monolayer exhibit contrasting Berry phase effect and optical circular dichroism. 
The above valley-dependent spin textures and optoelectronic properties also apply to SnP$_{2}$Se$_{6}$ few-layers and other MP$_{2}$X$_{6}$. 
These intriguing characteristics give full play to spin and valley degrees of freedom, and provide opportunities for designing advanced electronic devices.

{\color{blue}\textit{Acknowledgments}} -- 
We are supported by the National Natural Science Foundation of China (Nos. 12374044, 12004186, 11904173). 

%


\begin{thebibliography}{40}%
\makeatletter
\providecommand \@ifxundefined [1]{%
 \@ifx{#1\undefined}
}%
\providecommand \@ifnum [1]{%
 \ifnum #1\expandafter \@firstoftwo
 \else \expandafter \@secondoftwo
 \fi
}%
\providecommand \@ifx [1]{%
 \ifx #1\expandafter \@firstoftwo
 \else \expandafter \@secondoftwo
 \fi
}%
\providecommand \natexlab [1]{#1}%
\providecommand \enquote  [1]{``#1''}%
\providecommand \bibnamefont  [1]{#1}%
\providecommand \bibfnamefont [1]{#1}%
\providecommand \citenamefont [1]{#1}%
\providecommand \href@noop [0]{\@secondoftwo}%
\providecommand \href [0]{\begingroup \@sanitize@url \@href}%
\providecommand \@href[1]{\@@startlink{#1}\@@href}%
\providecommand \@@href[1]{\endgroup#1\@@endlink}%
\providecommand \@sanitize@url [0]{\catcode `\\12\catcode `\$12\catcode
  `\&12\catcode `\#12\catcode `\^12\catcode `\_12\catcode `\%12\relax}%
\providecommand \@@startlink[1]{}%
\providecommand \@@endlink[0]{}%
\providecommand \url  [0]{\begingroup\@sanitize@url \@url }%
\providecommand \@url [1]{\endgroup\@href {#1}{\urlprefix }}%
\providecommand \urlprefix  [0]{URL }%
\providecommand \Eprint [0]{\href }%
\providecommand \doibase [0]{http://dx.doi.org/}%
\providecommand \selectlanguage [0]{\@gobble}%
\providecommand \bibinfo  [0]{\@secondoftwo}%
\providecommand \bibfield  [0]{\@secondoftwo}%
\providecommand \translation [1]{[#1]}%
\providecommand \BibitemOpen [0]{}%
\providecommand \bibitemStop [0]{}%
\providecommand \bibitemNoStop [0]{.\EOS\space}%
\providecommand \EOS [0]{\spacefactor3000\relax}%
\providecommand \BibitemShut  [1]{\csname bibitem#1\endcsname}%
\let\auto@bib@innerbib\@empty
\bibitem [{\citenamefont {Manchon}\ \emph {et~al.}(2015)\citenamefont
  {Manchon}, \citenamefont {Koo}, \citenamefont {Nitta}, \citenamefont
  {Frolov},\ and\ \citenamefont {Duine}}]{Manchon2015}%
  \BibitemOpen
  \bibfield  {author} {\bibinfo {author} {\bibfnamefont {A.}~\bibnamefont
  {Manchon}}, \bibinfo {author} {\bibfnamefont {H.~C.}\ \bibnamefont {Koo}},
  \bibinfo {author} {\bibfnamefont {J.}~\bibnamefont {Nitta}}, \bibinfo
  {author} {\bibfnamefont {S.~M.}\ \bibnamefont {Frolov}}, \ and\ \bibinfo
  {author} {\bibfnamefont {R.~A.}\ \bibnamefont {Duine}},\ }\href {\doibase
  10.1038/nmat4360} {\bibfield  {journal} {\bibinfo  {journal} {Nature
  Materials}\ }\textbf {\bibinfo {volume} {14}},\ \bibinfo {pages} {871}
  (\bibinfo {year} {2015})}\BibitemShut {NoStop}%
\bibitem [{\citenamefont {Chen}\ \emph {et~al.}(2021)\citenamefont {Chen},
  \citenamefont {Wu}, \citenamefont {Hu},\ and\ \citenamefont
  {Yang}}]{Chen2021}%
  \BibitemOpen
  \bibfield  {author} {\bibinfo {author} {\bibfnamefont {J.}~\bibnamefont
  {Chen}}, \bibinfo {author} {\bibfnamefont {K.}~\bibnamefont {Wu}}, \bibinfo
  {author} {\bibfnamefont {W.}~\bibnamefont {Hu}}, \ and\ \bibinfo {author}
  {\bibfnamefont {J.}~\bibnamefont {Yang}},\ }\href {\doibase
  10.1021/acs.jpclett.1c03662} {\bibfield  {journal} {\bibinfo  {journal} {J.
  Phys. Chem. Lett.}\ }\textbf {\bibinfo {volume} {12}},\ \bibinfo {pages}
  {12256} (\bibinfo {year} {2021})}\BibitemShut {NoStop}%
\bibitem [{\citenamefont {Vasko}(1979)}]{Vasko1979}%
  \BibitemOpen
  \bibfield  {author} {\bibinfo {author} {\bibfnamefont {F.~T.}\ \bibnamefont
  {Vasko}},\ }\href@noop {} {\bibfield  {journal} {\bibinfo  {journal} {JETP
  Lett.}\ }\textbf {\bibinfo {volume} {30}},\ \bibinfo {pages} {541} (\bibinfo
  {year} {1979})}\BibitemShut {NoStop}%
\bibitem [{\citenamefont {Bychkov}\ and\ \citenamefont
  {Rashba}(1984)}]{Bychkov1984}%
  \BibitemOpen
  \bibfield  {author} {\bibinfo {author} {\bibfnamefont {Y.~A.}\ \bibnamefont
  {Bychkov}}\ and\ \bibinfo {author} {\bibfnamefont {E.~I.}\ \bibnamefont
  {Rashba}},\ }\href@noop {} {\bibfield  {journal} {\bibinfo  {journal} {JETP
  Lett.}\ }\textbf {\bibinfo {volume} {39}},\ \bibinfo {pages} {78} (\bibinfo
  {year} {1984})}\BibitemShut {NoStop}%
\bibitem [{\citenamefont {Nitta}\ \emph {et~al.}(1997)\citenamefont {Nitta},
  \citenamefont {Akazaki}, \citenamefont {Takayanagi},\ and\ \citenamefont
  {Enoki}}]{Nitta1997}%
  \BibitemOpen
  \bibfield  {author} {\bibinfo {author} {\bibfnamefont {J.}~\bibnamefont
  {Nitta}}, \bibinfo {author} {\bibfnamefont {T.}~\bibnamefont {Akazaki}},
  \bibinfo {author} {\bibfnamefont {H.}~\bibnamefont {Takayanagi}}, \ and\
  \bibinfo {author} {\bibfnamefont {T.}~\bibnamefont {Enoki}},\ }\href
  {\doibase 10.1103/PhysRevLett.78.1335} {\bibfield  {journal} {\bibinfo
  {journal} {Phys. Rev. Lett.}\ }\textbf {\bibinfo {volume} {78}},\ \bibinfo
  {pages} {1335} (\bibinfo {year} {1997})}\BibitemShut {NoStop}%
\bibitem [{\citenamefont {Dresselhaus}(1955)}]{Dresselhaus1955}%
  \BibitemOpen
  \bibfield  {author} {\bibinfo {author} {\bibfnamefont {G.}~\bibnamefont
  {Dresselhaus}},\ }\href {\doibase 10.1103/PhysRev.100.580} {\bibfield
  {journal} {\bibinfo  {journal} {Phys. Rev.}\ }\textbf {\bibinfo {volume}
  {100}},\ \bibinfo {pages} {580} (\bibinfo {year} {1955})}\BibitemShut
  {NoStop}%
\bibitem [{\citenamefont {Zhu}\ \emph {et~al.}(2011)\citenamefont {Zhu},
  \citenamefont {Cheng},\ and\ \citenamefont {Schwingenschl\"ogl}}]{Zhu2011}%
  \BibitemOpen
  \bibfield  {author} {\bibinfo {author} {\bibfnamefont {Z.~Y.}\ \bibnamefont
  {Zhu}}, \bibinfo {author} {\bibfnamefont {Y.~C.}\ \bibnamefont {Cheng}}, \
  and\ \bibinfo {author} {\bibfnamefont {U.}~\bibnamefont
  {Schwingenschl\"ogl}},\ }\href {\doibase 10.1103/PhysRevB.84.153402}
  {\bibfield  {journal} {\bibinfo  {journal} {Phys. Rev. B}\ }\textbf {\bibinfo
  {volume} {84}},\ \bibinfo {pages} {153402} (\bibinfo {year}
  {2011})}\BibitemShut {NoStop}%
\bibitem [{\citenamefont {Xiao}\ \emph {et~al.}(2012)\citenamefont {Xiao},
  \citenamefont {Liu}, \citenamefont {Feng}, \citenamefont {Xu},\ and\
  \citenamefont {Yao}}]{Xiao2012}%
  \BibitemOpen
  \bibfield  {author} {\bibinfo {author} {\bibfnamefont {D.}~\bibnamefont
  {Xiao}}, \bibinfo {author} {\bibfnamefont {G.-B.}\ \bibnamefont {Liu}},
  \bibinfo {author} {\bibfnamefont {W.}~\bibnamefont {Feng}}, \bibinfo {author}
  {\bibfnamefont {X.}~\bibnamefont {Xu}}, \ and\ \bibinfo {author}
  {\bibfnamefont {W.}~\bibnamefont {Yao}},\ }\href {\doibase
  10.1103/PhysRevLett.108.196802} {\bibfield  {journal} {\bibinfo  {journal}
  {Phys. Rev. Lett.}\ }\textbf {\bibinfo {volume} {108}},\ \bibinfo {pages}
  {196802} (\bibinfo {year} {2012})}\BibitemShut {NoStop}%
\bibitem [{\citenamefont {Hirayama}\ \emph {et~al.}(2015)\citenamefont
  {Hirayama}, \citenamefont {Okugawa}, \citenamefont {Ishibashi}, \citenamefont
  {Murakami},\ and\ \citenamefont {Miyake}}]{Hirayama2015}%
  \BibitemOpen
  \bibfield  {author} {\bibinfo {author} {\bibfnamefont {M.}~\bibnamefont
  {Hirayama}}, \bibinfo {author} {\bibfnamefont {R.}~\bibnamefont {Okugawa}},
  \bibinfo {author} {\bibfnamefont {S.}~\bibnamefont {Ishibashi}}, \bibinfo
  {author} {\bibfnamefont {S.}~\bibnamefont {Murakami}}, \ and\ \bibinfo
  {author} {\bibfnamefont {T.}~\bibnamefont {Miyake}},\ }\href {\doibase
  10.1103/PhysRevLett.114.206401} {\bibfield  {journal} {\bibinfo  {journal}
  {Phys. Rev. Lett.}\ }\textbf {\bibinfo {volume} {114}},\ \bibinfo {pages}
  {206401} (\bibinfo {year} {2015})}\BibitemShut {NoStop}%
\bibitem [{\citenamefont {Sakano}\ \emph {et~al.}(2020)\citenamefont {Sakano},
  \citenamefont {Hirayama}, \citenamefont {Takahashi}, \citenamefont {Akebi},
  \citenamefont {Nakayama}, \citenamefont {Kuroda}, \citenamefont {Taguchi},
  \citenamefont {Yoshikawa}, \citenamefont {Miyamoto}, \citenamefont {Okuda},
  \citenamefont {Ono}, \citenamefont {Kumigashira}, \citenamefont {Ideue},
  \citenamefont {Iwasa}, \citenamefont {Mitsuishi}, \citenamefont {Ishizaka},
  \citenamefont {Shin}, \citenamefont {Miyake}, \citenamefont {Murakami},
  \citenamefont {Sasagawa},\ and\ \citenamefont {Kondo}}]{Sakano2020}%
  \BibitemOpen
  \bibfield  {author} {\bibinfo {author} {\bibfnamefont {M.}~\bibnamefont
  {Sakano}}, \bibinfo {author} {\bibfnamefont {M.}~\bibnamefont {Hirayama}},
  \bibinfo {author} {\bibfnamefont {T.}~\bibnamefont {Takahashi}}, \bibinfo
  {author} {\bibfnamefont {S.}~\bibnamefont {Akebi}}, \bibinfo {author}
  {\bibfnamefont {M.}~\bibnamefont {Nakayama}}, \bibinfo {author}
  {\bibfnamefont {K.}~\bibnamefont {Kuroda}}, \bibinfo {author} {\bibfnamefont
  {K.}~\bibnamefont {Taguchi}}, \bibinfo {author} {\bibfnamefont
  {T.}~\bibnamefont {Yoshikawa}}, \bibinfo {author} {\bibfnamefont
  {K.}~\bibnamefont {Miyamoto}}, \bibinfo {author} {\bibfnamefont
  {T.}~\bibnamefont {Okuda}}, \bibinfo {author} {\bibfnamefont
  {K.}~\bibnamefont {Ono}}, \bibinfo {author} {\bibfnamefont {H.}~\bibnamefont
  {Kumigashira}}, \bibinfo {author} {\bibfnamefont {T.}~\bibnamefont {Ideue}},
  \bibinfo {author} {\bibfnamefont {Y.}~\bibnamefont {Iwasa}}, \bibinfo
  {author} {\bibfnamefont {N.}~\bibnamefont {Mitsuishi}}, \bibinfo {author}
  {\bibfnamefont {K.}~\bibnamefont {Ishizaka}}, \bibinfo {author}
  {\bibfnamefont {S.}~\bibnamefont {Shin}}, \bibinfo {author} {\bibfnamefont
  {T.}~\bibnamefont {Miyake}}, \bibinfo {author} {\bibfnamefont
  {S.}~\bibnamefont {Murakami}}, \bibinfo {author} {\bibfnamefont
  {T.}~\bibnamefont {Sasagawa}}, \ and\ \bibinfo {author} {\bibfnamefont
  {T.}~\bibnamefont {Kondo}},\ }\href {\doibase 10.1103/PhysRevLett.124.136404}
  {\bibfield  {journal} {\bibinfo  {journal} {Phys. Rev. Lett.}\ }\textbf
  {\bibinfo {volume} {124}},\ \bibinfo {pages} {136404} (\bibinfo {year}
  {2020})}\BibitemShut {NoStop}%
\bibitem [{\citenamefont {Gatti}\ \emph {et~al.}(2020)\citenamefont {Gatti},
  \citenamefont {Gos\'albez-Mart\'{\i}nez}, \citenamefont {Tsirkin},
  \citenamefont {Fanciulli}, \citenamefont {Puppin}, \citenamefont
  {Polishchuk}, \citenamefont {Moser}, \citenamefont {Testa}, \citenamefont
  {Martino}, \citenamefont {Roth}, \citenamefont {Bugnon}, \citenamefont
  {Moreschini}, \citenamefont {Bostwick}, \citenamefont {Jozwiak},
  \citenamefont {Rotenberg}, \citenamefont {Di~Santo}, \citenamefont
  {Petaccia}, \citenamefont {Vobornik}, \citenamefont {Fujii}, \citenamefont
  {Wong}, \citenamefont {Jariwala}, \citenamefont {Atwater}, \citenamefont
  {R\o{}nnow}, \citenamefont {Chergui}, \citenamefont {Yazyev}, \citenamefont
  {Grioni},\ and\ \citenamefont {Crepaldi}}]{Gatti2020}%
  \BibitemOpen
  \bibfield  {author} {\bibinfo {author} {\bibfnamefont {G.}~\bibnamefont
  {Gatti}}, \bibinfo {author} {\bibfnamefont {D.}~\bibnamefont
  {Gos\'albez-Mart\'{\i}nez}}, \bibinfo {author} {\bibfnamefont {S.~S.}\
  \bibnamefont {Tsirkin}}, \bibinfo {author} {\bibfnamefont {M.}~\bibnamefont
  {Fanciulli}}, \bibinfo {author} {\bibfnamefont {M.}~\bibnamefont {Puppin}},
  \bibinfo {author} {\bibfnamefont {S.}~\bibnamefont {Polishchuk}}, \bibinfo
  {author} {\bibfnamefont {S.}~\bibnamefont {Moser}}, \bibinfo {author}
  {\bibfnamefont {L.}~\bibnamefont {Testa}}, \bibinfo {author} {\bibfnamefont
  {E.}~\bibnamefont {Martino}}, \bibinfo {author} {\bibfnamefont
  {S.}~\bibnamefont {Roth}}, \bibinfo {author} {\bibfnamefont {P.}~\bibnamefont
  {Bugnon}}, \bibinfo {author} {\bibfnamefont {L.}~\bibnamefont {Moreschini}},
  \bibinfo {author} {\bibfnamefont {A.}~\bibnamefont {Bostwick}}, \bibinfo
  {author} {\bibfnamefont {C.}~\bibnamefont {Jozwiak}}, \bibinfo {author}
  {\bibfnamefont {E.}~\bibnamefont {Rotenberg}}, \bibinfo {author}
  {\bibfnamefont {G.}~\bibnamefont {Di~Santo}}, \bibinfo {author}
  {\bibfnamefont {L.}~\bibnamefont {Petaccia}}, \bibinfo {author}
  {\bibfnamefont {I.}~\bibnamefont {Vobornik}}, \bibinfo {author}
  {\bibfnamefont {J.}~\bibnamefont {Fujii}}, \bibinfo {author} {\bibfnamefont
  {J.}~\bibnamefont {Wong}}, \bibinfo {author} {\bibfnamefont {D.}~\bibnamefont
  {Jariwala}}, \bibinfo {author} {\bibfnamefont {H.~A.}\ \bibnamefont
  {Atwater}}, \bibinfo {author} {\bibfnamefont {H.~M.}\ \bibnamefont
  {R\o{}nnow}}, \bibinfo {author} {\bibfnamefont {M.}~\bibnamefont {Chergui}},
  \bibinfo {author} {\bibfnamefont {O.~V.}\ \bibnamefont {Yazyev}}, \bibinfo
  {author} {\bibfnamefont {M.}~\bibnamefont {Grioni}}, \ and\ \bibinfo {author}
  {\bibfnamefont {A.}~\bibnamefont {Crepaldi}},\ }\href {\doibase
  10.1103/PhysRevLett.125.216402} {\bibfield  {journal} {\bibinfo  {journal}
  {Phys. Rev. Lett.}\ }\textbf {\bibinfo {volume} {125}},\ \bibinfo {pages}
  {216402} (\bibinfo {year} {2020})}\BibitemShut {NoStop}%
\bibitem [{\citenamefont {Krieger}\ \emph {et~al.}(2024)\citenamefont
  {Krieger}, \citenamefont {Stolz}, \citenamefont {Robredo}, \citenamefont
  {Manna}, \citenamefont {McFarlane}, \citenamefont {Date}, \citenamefont
  {Pal}, \citenamefont {Yang}, \citenamefont {B.~Guedes}, \citenamefont {Dil},
  \citenamefont {Polley}, \citenamefont {Leandersson}, \citenamefont {Shekhar},
  \citenamefont {Borrmann}, \citenamefont {Yang}, \citenamefont {Lin},
  \citenamefont {Strocov}, \citenamefont {Caputo}, \citenamefont {Watson},
  \citenamefont {Kim}, \citenamefont {Cacho}, \citenamefont {Mazzola},
  \citenamefont {Fujii}, \citenamefont {Vobornik}, \citenamefont {Parkin},
  \citenamefont {Bradlyn}, \citenamefont {Felser}, \citenamefont {Vergniory},\
  and\ \citenamefont {Schröter}}]{Krieger2024}%
  \BibitemOpen
  \bibfield  {author} {\bibinfo {author} {\bibfnamefont {J.~A.}\ \bibnamefont
  {Krieger}}, \bibinfo {author} {\bibfnamefont {S.}~\bibnamefont {Stolz}},
  \bibinfo {author} {\bibfnamefont {I.}~\bibnamefont {Robredo}}, \bibinfo
  {author} {\bibfnamefont {K.}~\bibnamefont {Manna}}, \bibinfo {author}
  {\bibfnamefont {E.~C.}\ \bibnamefont {McFarlane}}, \bibinfo {author}
  {\bibfnamefont {M.}~\bibnamefont {Date}}, \bibinfo {author} {\bibfnamefont
  {B.}~\bibnamefont {Pal}}, \bibinfo {author} {\bibfnamefont {J.}~\bibnamefont
  {Yang}}, \bibinfo {author} {\bibfnamefont {E.}~\bibnamefont {B.~Guedes}},
  \bibinfo {author} {\bibfnamefont {J.~H.}\ \bibnamefont {Dil}}, \bibinfo
  {author} {\bibfnamefont {C.~M.}\ \bibnamefont {Polley}}, \bibinfo {author}
  {\bibfnamefont {M.}~\bibnamefont {Leandersson}}, \bibinfo {author}
  {\bibfnamefont {C.}~\bibnamefont {Shekhar}}, \bibinfo {author} {\bibfnamefont
  {H.}~\bibnamefont {Borrmann}}, \bibinfo {author} {\bibfnamefont
  {Q.}~\bibnamefont {Yang}}, \bibinfo {author} {\bibfnamefont {M.}~\bibnamefont
  {Lin}}, \bibinfo {author} {\bibfnamefont {V.~N.}\ \bibnamefont {Strocov}},
  \bibinfo {author} {\bibfnamefont {M.}~\bibnamefont {Caputo}}, \bibinfo
  {author} {\bibfnamefont {M.~D.}\ \bibnamefont {Watson}}, \bibinfo {author}
  {\bibfnamefont {T.~K.}\ \bibnamefont {Kim}}, \bibinfo {author} {\bibfnamefont
  {C.}~\bibnamefont {Cacho}}, \bibinfo {author} {\bibfnamefont
  {F.}~\bibnamefont {Mazzola}}, \bibinfo {author} {\bibfnamefont
  {J.}~\bibnamefont {Fujii}}, \bibinfo {author} {\bibfnamefont
  {I.}~\bibnamefont {Vobornik}}, \bibinfo {author} {\bibfnamefont {S.~S.~P.}\
  \bibnamefont {Parkin}}, \bibinfo {author} {\bibfnamefont {B.}~\bibnamefont
  {Bradlyn}}, \bibinfo {author} {\bibfnamefont {C.}~\bibnamefont {Felser}},
  \bibinfo {author} {\bibfnamefont {M.~G.}\ \bibnamefont {Vergniory}}, \ and\
  \bibinfo {author} {\bibfnamefont {N.~B.~M.}\ \bibnamefont {Schröter}},\
  }\href {\doibase 10.1038/s41467-024-47976-0} {\bibfield  {journal} {\bibinfo
  {journal} {Nature Communications}\ }\textbf {\bibinfo {volume} {15}},\
  \bibinfo {pages} {3720} (\bibinfo {year} {2024})}\BibitemShut {NoStop}%
\bibitem [{\citenamefont {Yuan}\ \emph {et~al.}(2014)\citenamefont {Yuan},
  \citenamefont {Mak},\ and\ \citenamefont {Law}}]{Yuan2014}%
  \BibitemOpen
  \bibfield  {author} {\bibinfo {author} {\bibfnamefont {N.~F.~Q.}\
  \bibnamefont {Yuan}}, \bibinfo {author} {\bibfnamefont {K.~F.}\ \bibnamefont
  {Mak}}, \ and\ \bibinfo {author} {\bibfnamefont {K.~T.}\ \bibnamefont
  {Law}},\ }\href {\doibase 10.1103/PhysRevLett.113.097001} {\bibfield
  {journal} {\bibinfo  {journal} {Phys. Rev. Lett.}\ }\textbf {\bibinfo
  {volume} {113}},\ \bibinfo {pages} {097001} (\bibinfo {year}
  {2014})}\BibitemShut {NoStop}%
\bibitem [{\citenamefont {Lu}\ \emph {et~al.}(2015)\citenamefont {Lu},
  \citenamefont {Zheliuk}, \citenamefont {Leermakers}, \citenamefont {Yuan},
  \citenamefont {Zeitler}, \citenamefont {Law},\ and\ \citenamefont
  {Ye}}]{Lu2015}%
  \BibitemOpen
  \bibfield  {author} {\bibinfo {author} {\bibfnamefont {J.~M.}\ \bibnamefont
  {Lu}}, \bibinfo {author} {\bibfnamefont {O.}~\bibnamefont {Zheliuk}},
  \bibinfo {author} {\bibfnamefont {I.}~\bibnamefont {Leermakers}}, \bibinfo
  {author} {\bibfnamefont {N.~F.~Q.}\ \bibnamefont {Yuan}}, \bibinfo {author}
  {\bibfnamefont {U.}~\bibnamefont {Zeitler}}, \bibinfo {author} {\bibfnamefont
  {K.~T.}\ \bibnamefont {Law}}, \ and\ \bibinfo {author} {\bibfnamefont
  {J.~T.}\ \bibnamefont {Ye}},\ }\href {\doibase 10.1126/science.aab2277}
  {\bibfield  {journal} {\bibinfo  {journal} {Science}\ }\textbf {\bibinfo
  {volume} {350}},\ \bibinfo {pages} {1353} (\bibinfo {year}
  {2015})}\BibitemShut {NoStop}%
\bibitem [{\citenamefont {Wang}\ \emph {et~al.}(2019)\citenamefont {Wang},
  \citenamefont {Lian}, \citenamefont {Guo}, \citenamefont {Mao}, \citenamefont
  {Zhang}, \citenamefont {Zhang}, \citenamefont {Gu}, \citenamefont {Xu},\ and\
  \citenamefont {Duan}}]{Wang2019}%
  \BibitemOpen
  \bibfield  {author} {\bibinfo {author} {\bibfnamefont {C.}~\bibnamefont
  {Wang}}, \bibinfo {author} {\bibfnamefont {B.}~\bibnamefont {Lian}}, \bibinfo
  {author} {\bibfnamefont {X.}~\bibnamefont {Guo}}, \bibinfo {author}
  {\bibfnamefont {J.}~\bibnamefont {Mao}}, \bibinfo {author} {\bibfnamefont
  {Z.}~\bibnamefont {Zhang}}, \bibinfo {author} {\bibfnamefont
  {D.}~\bibnamefont {Zhang}}, \bibinfo {author} {\bibfnamefont {B.-L.}\
  \bibnamefont {Gu}}, \bibinfo {author} {\bibfnamefont {Y.}~\bibnamefont {Xu}},
  \ and\ \bibinfo {author} {\bibfnamefont {W.}~\bibnamefont {Duan}},\ }\href
  {\doibase 10.1103/PhysRevLett.123.126402} {\bibfield  {journal} {\bibinfo
  {journal} {Phys. Rev. Lett.}\ }\textbf {\bibinfo {volume} {123}},\ \bibinfo
  {pages} {126402} (\bibinfo {year} {2019})}\BibitemShut {NoStop}%
\bibitem [{\citenamefont {Falson}\ \emph {et~al.}(2020)\citenamefont {Falson},
  \citenamefont {Xu}, \citenamefont {Liao}, \citenamefont {Zang}, \citenamefont
  {Zhu}, \citenamefont {Wang}, \citenamefont {Zhang}, \citenamefont {Liu},
  \citenamefont {Duan}, \citenamefont {He}, \citenamefont {Liu}, \citenamefont
  {Smet}, \citenamefont {Zhang},\ and\ \citenamefont {Xue}}]{Falson2020}%
  \BibitemOpen
  \bibfield  {author} {\bibinfo {author} {\bibfnamefont {J.}~\bibnamefont
  {Falson}}, \bibinfo {author} {\bibfnamefont {Y.}~\bibnamefont {Xu}}, \bibinfo
  {author} {\bibfnamefont {M.}~\bibnamefont {Liao}}, \bibinfo {author}
  {\bibfnamefont {Y.}~\bibnamefont {Zang}}, \bibinfo {author} {\bibfnamefont
  {K.}~\bibnamefont {Zhu}}, \bibinfo {author} {\bibfnamefont {C.}~\bibnamefont
  {Wang}}, \bibinfo {author} {\bibfnamefont {Z.}~\bibnamefont {Zhang}},
  \bibinfo {author} {\bibfnamefont {H.}~\bibnamefont {Liu}}, \bibinfo {author}
  {\bibfnamefont {W.}~\bibnamefont {Duan}}, \bibinfo {author} {\bibfnamefont
  {K.}~\bibnamefont {He}}, \bibinfo {author} {\bibfnamefont {H.}~\bibnamefont
  {Liu}}, \bibinfo {author} {\bibfnamefont {J.~H.}\ \bibnamefont {Smet}},
  \bibinfo {author} {\bibfnamefont {D.}~\bibnamefont {Zhang}}, \ and\ \bibinfo
  {author} {\bibfnamefont {Q.-K.}\ \bibnamefont {Xue}},\ }\href {\doibase
  10.1126/science.aax3873} {\bibfield  {journal} {\bibinfo  {journal}
  {Science}\ }\textbf {\bibinfo {volume} {367}},\ \bibinfo {pages} {1454}
  (\bibinfo {year} {2020})}\BibitemShut {NoStop}%
\bibitem [{\citenamefont {Liu}\ \emph {et~al.}(2020)\citenamefont {Liu},
  \citenamefont {Xu}, \citenamefont {Sun}, \citenamefont {Liu}, \citenamefont
  {Liu}, \citenamefont {Wang}, \citenamefont {Zhang}, \citenamefont {Gu},
  \citenamefont {Tang}, \citenamefont {Ding}, \citenamefont {Liu},
  \citenamefont {Yao}, \citenamefont {Lin}, \citenamefont {Wang}, \citenamefont
  {Xue},\ and\ \citenamefont {Wang}}]{Liu2020}%
  \BibitemOpen
  \bibfield  {author} {\bibinfo {author} {\bibfnamefont {Y.}~\bibnamefont
  {Liu}}, \bibinfo {author} {\bibfnamefont {Y.}~\bibnamefont {Xu}}, \bibinfo
  {author} {\bibfnamefont {J.}~\bibnamefont {Sun}}, \bibinfo {author}
  {\bibfnamefont {C.}~\bibnamefont {Liu}}, \bibinfo {author} {\bibfnamefont
  {Y.}~\bibnamefont {Liu}}, \bibinfo {author} {\bibfnamefont {C.}~\bibnamefont
  {Wang}}, \bibinfo {author} {\bibfnamefont {Z.}~\bibnamefont {Zhang}},
  \bibinfo {author} {\bibfnamefont {K.}~\bibnamefont {Gu}}, \bibinfo {author}
  {\bibfnamefont {Y.}~\bibnamefont {Tang}}, \bibinfo {author} {\bibfnamefont
  {C.}~\bibnamefont {Ding}}, \bibinfo {author} {\bibfnamefont {H.}~\bibnamefont
  {Liu}}, \bibinfo {author} {\bibfnamefont {H.}~\bibnamefont {Yao}}, \bibinfo
  {author} {\bibfnamefont {X.}~\bibnamefont {Lin}}, \bibinfo {author}
  {\bibfnamefont {L.}~\bibnamefont {Wang}}, \bibinfo {author} {\bibfnamefont
  {Q.-K.}\ \bibnamefont {Xue}}, \ and\ \bibinfo {author} {\bibfnamefont
  {J.}~\bibnamefont {Wang}},\ }\href {\doibase 10.1021/acs.nanolett.0c01356}
  {\bibfield  {journal} {\bibinfo  {journal} {Nano Lett.}\ }\textbf {\bibinfo
  {volume} {20}},\ \bibinfo {pages} {5728} (\bibinfo {year} {2020})},\ \bibinfo
  {note} {publisher: American Chemical Society}\BibitemShut {NoStop}%
\bibitem [{\citenamefont {Xie}\ \emph {et~al.}(2020)\citenamefont {Xie},
  \citenamefont {Zhou},\ and\ \citenamefont {Law}}]{Xie2020}%
  \BibitemOpen
  \bibfield  {author} {\bibinfo {author} {\bibfnamefont {Y.-M.}\ \bibnamefont
  {Xie}}, \bibinfo {author} {\bibfnamefont {B.~T.}\ \bibnamefont {Zhou}}, \
  and\ \bibinfo {author} {\bibfnamefont {K.~T.}\ \bibnamefont {Law}},\ }\href
  {\doibase 10.1103/PhysRevLett.125.107001} {\bibfield  {journal} {\bibinfo
  {journal} {Phys. Rev. Lett.}\ }\textbf {\bibinfo {volume} {125}},\ \bibinfo
  {pages} {107001} (\bibinfo {year} {2020})}\BibitemShut {NoStop}%
\bibitem [{\citenamefont {Zhang}\ \emph {et~al.}(2023)\citenamefont {Zhang},
  \citenamefont {Xie}, \citenamefont {Fang}, \citenamefont {Zhang},
  \citenamefont {Xu}, \citenamefont {Zou}, \citenamefont {Leng}, \citenamefont
  {Gao}, \citenamefont {Zhang}, \citenamefont {Ai}, \citenamefont {Zhang},
  \citenamefont {Jia}, \citenamefont {Liu}, \citenamefont {Yan}, \citenamefont
  {Zhao}, \citenamefont {Haigh}, \citenamefont {Kou}, \citenamefont {Yang},
  \citenamefont {Huang}, \citenamefont {Law}, \citenamefont {Xiu},\ and\
  \citenamefont {Dong}}]{ZhangE2023}%
  \BibitemOpen
  \bibfield  {author} {\bibinfo {author} {\bibfnamefont {E.}~\bibnamefont
  {Zhang}}, \bibinfo {author} {\bibfnamefont {Y.-M.}\ \bibnamefont {Xie}},
  \bibinfo {author} {\bibfnamefont {Y.}~\bibnamefont {Fang}}, \bibinfo {author}
  {\bibfnamefont {J.}~\bibnamefont {Zhang}}, \bibinfo {author} {\bibfnamefont
  {X.}~\bibnamefont {Xu}}, \bibinfo {author} {\bibfnamefont {Y.-C.}\
  \bibnamefont {Zou}}, \bibinfo {author} {\bibfnamefont {P.}~\bibnamefont
  {Leng}}, \bibinfo {author} {\bibfnamefont {X.-J.}\ \bibnamefont {Gao}},
  \bibinfo {author} {\bibfnamefont {Y.}~\bibnamefont {Zhang}}, \bibinfo
  {author} {\bibfnamefont {L.}~\bibnamefont {Ai}}, \bibinfo {author}
  {\bibfnamefont {Y.}~\bibnamefont {Zhang}}, \bibinfo {author} {\bibfnamefont
  {Z.}~\bibnamefont {Jia}}, \bibinfo {author} {\bibfnamefont {S.}~\bibnamefont
  {Liu}}, \bibinfo {author} {\bibfnamefont {J.}~\bibnamefont {Yan}}, \bibinfo
  {author} {\bibfnamefont {W.}~\bibnamefont {Zhao}}, \bibinfo {author}
  {\bibfnamefont {S.~J.}\ \bibnamefont {Haigh}}, \bibinfo {author}
  {\bibfnamefont {X.}~\bibnamefont {Kou}}, \bibinfo {author} {\bibfnamefont
  {J.}~\bibnamefont {Yang}}, \bibinfo {author} {\bibfnamefont {F.}~\bibnamefont
  {Huang}}, \bibinfo {author} {\bibfnamefont {K.~T.}\ \bibnamefont {Law}},
  \bibinfo {author} {\bibfnamefont {F.}~\bibnamefont {Xiu}}, \ and\ \bibinfo
  {author} {\bibfnamefont {S.}~\bibnamefont {Dong}},\ }\href {\doibase
  10.1038/s41567-022-01812-8} {\bibfield  {journal} {\bibinfo  {journal}
  {Nature Physics}\ }\textbf {\bibinfo {volume} {19}},\ \bibinfo {pages} {106}
  (\bibinfo {year} {2023})}\BibitemShut {NoStop}%
\bibitem [{\citenamefont {Fischer}\ \emph {et~al.}(2023)\citenamefont
  {Fischer}, \citenamefont {Sigrist}, \citenamefont {Agterberg},\ and\
  \citenamefont {Yanase}}]{Fischer2023}%
  \BibitemOpen
  \bibfield  {author} {\bibinfo {author} {\bibfnamefont {M.~H.}\ \bibnamefont
  {Fischer}}, \bibinfo {author} {\bibfnamefont {M.}~\bibnamefont {Sigrist}},
  \bibinfo {author} {\bibfnamefont {D.~F.}\ \bibnamefont {Agterberg}}, \ and\
  \bibinfo {author} {\bibfnamefont {Y.}~\bibnamefont {Yanase}},\ }\href
  {\doibase https://doi.org/10.1146/annurev-conmatphys-040521-042511}
  {\bibfield  {journal} {\bibinfo  {journal} {Annual Review of Condensed Matter
  Physics}\ }\textbf {\bibinfo {volume} {14}},\ \bibinfo {pages} {153}
  (\bibinfo {year} {2023})}\BibitemShut {NoStop}%
\bibitem [{\citenamefont {Shao}\ \emph {et~al.}(2016)\citenamefont {Shao},
  \citenamefont {Yu}, \citenamefont {Lan}, \citenamefont {Shi}, \citenamefont
  {Li}, \citenamefont {Zheng}, \citenamefont {Zhu}, \citenamefont {Li},
  \citenamefont {Amiri},\ and\ \citenamefont {Wang}}]{Shao2016}%
  \BibitemOpen
  \bibfield  {author} {\bibinfo {author} {\bibfnamefont {Q.}~\bibnamefont
  {Shao}}, \bibinfo {author} {\bibfnamefont {G.}~\bibnamefont {Yu}}, \bibinfo
  {author} {\bibfnamefont {Y.-W.}\ \bibnamefont {Lan}}, \bibinfo {author}
  {\bibfnamefont {Y.}~\bibnamefont {Shi}}, \bibinfo {author} {\bibfnamefont
  {M.-Y.}\ \bibnamefont {Li}}, \bibinfo {author} {\bibfnamefont
  {C.}~\bibnamefont {Zheng}}, \bibinfo {author} {\bibfnamefont
  {X.}~\bibnamefont {Zhu}}, \bibinfo {author} {\bibfnamefont {L.-J.}\
  \bibnamefont {Li}}, \bibinfo {author} {\bibfnamefont {P.~K.}\ \bibnamefont
  {Amiri}}, \ and\ \bibinfo {author} {\bibfnamefont {K.~L.}\ \bibnamefont
  {Wang}},\ }\href {\doibase 10.1021/acs.nanolett.6b03300} {\bibfield
  {journal} {\bibinfo  {journal} {Nano Lett.}\ }\textbf {\bibinfo {volume}
  {16}},\ \bibinfo {pages} {7514} (\bibinfo {year} {2016})},\ \bibinfo {note}
  {publisher: American Chemical Society}\BibitemShut {NoStop}%
\bibitem [{\citenamefont {Li}\ \emph {et~al.}(2020{\natexlab{a}})\citenamefont
  {Li}, \citenamefont {Chen},\ and\ \citenamefont {Niu}}]{Li2020}%
  \BibitemOpen
  \bibfield  {author} {\bibinfo {author} {\bibfnamefont {X.}~\bibnamefont
  {Li}}, \bibinfo {author} {\bibfnamefont {H.}~\bibnamefont {Chen}}, \ and\
  \bibinfo {author} {\bibfnamefont {Q.}~\bibnamefont {Niu}},\ }\href {\doibase
  10.1073/pnas.1912472117} {\bibfield  {journal} {\bibinfo  {journal}
  {Proceedings of the National Academy of Sciences}\ }\textbf {\bibinfo
  {volume} {117}},\ \bibinfo {pages} {16749} (\bibinfo {year}
  {2020}{\natexlab{a}})}\BibitemShut {NoStop}%
\bibitem [{\citenamefont {Kang}\ \emph {et~al.}(2021)\citenamefont {Kang},
  \citenamefont {Choi}, \citenamefont {Jeong}, \citenamefont {Park},
  \citenamefont {Park}, \citenamefont {Kim}, \citenamefont {Lee}, \citenamefont
  {Kim}, \citenamefont {Kim}, \citenamefont {Oh}, \citenamefont {Viet},
  \citenamefont {Jeong}, \citenamefont {Yuk}, \citenamefont {Park},
  \citenamefont {Lee},\ and\ \citenamefont {Park}}]{Kang2021}%
  \BibitemOpen
  \bibfield  {author} {\bibinfo {author} {\bibfnamefont {M.-G.}\ \bibnamefont
  {Kang}}, \bibinfo {author} {\bibfnamefont {J.-G.}\ \bibnamefont {Choi}},
  \bibinfo {author} {\bibfnamefont {J.}~\bibnamefont {Jeong}}, \bibinfo
  {author} {\bibfnamefont {J.~Y.}\ \bibnamefont {Park}}, \bibinfo {author}
  {\bibfnamefont {H.-J.}\ \bibnamefont {Park}}, \bibinfo {author}
  {\bibfnamefont {T.}~\bibnamefont {Kim}}, \bibinfo {author} {\bibfnamefont
  {T.}~\bibnamefont {Lee}}, \bibinfo {author} {\bibfnamefont {K.-J.}\
  \bibnamefont {Kim}}, \bibinfo {author} {\bibfnamefont {K.-W.}\ \bibnamefont
  {Kim}}, \bibinfo {author} {\bibfnamefont {J.~H.}\ \bibnamefont {Oh}},
  \bibinfo {author} {\bibfnamefont {D.~D.}\ \bibnamefont {Viet}}, \bibinfo
  {author} {\bibfnamefont {J.-R.}\ \bibnamefont {Jeong}}, \bibinfo {author}
  {\bibfnamefont {J.~M.}\ \bibnamefont {Yuk}}, \bibinfo {author} {\bibfnamefont
  {J.}~\bibnamefont {Park}}, \bibinfo {author} {\bibfnamefont {K.-J.}\
  \bibnamefont {Lee}}, \ and\ \bibinfo {author} {\bibfnamefont {B.-G.}\
  \bibnamefont {Park}},\ }\href {\doibase 10.1038/s41467-021-27459-2}
  {\bibfield  {journal} {\bibinfo  {journal} {Nature Communications}\ }\textbf
  {\bibinfo {volume} {12}},\ \bibinfo {pages} {7111} (\bibinfo {year}
  {2021})}\BibitemShut {NoStop}%
\bibitem [{\citenamefont {He}\ \emph {et~al.}(2021)\citenamefont {He},
  \citenamefont {Xu},\ and\ \citenamefont {Law}}]{He2021}%
  \BibitemOpen
  \bibfield  {author} {\bibinfo {author} {\bibfnamefont {W.-Y.}\ \bibnamefont
  {He}}, \bibinfo {author} {\bibfnamefont {X.~Y.}\ \bibnamefont {Xu}}, \ and\
  \bibinfo {author} {\bibfnamefont {K.~T.}\ \bibnamefont {Law}},\ }\href
  {\doibase 10.1038/s42005-021-00564-w} {\bibfield  {journal} {\bibinfo
  {journal} {Communications Physics}\ }\textbf {\bibinfo {volume} {4}},\
  \bibinfo {pages} {66} (\bibinfo {year} {2021})}\BibitemShut {NoStop}%
\bibitem [{\citenamefont {Burkov}\ \emph {et~al.}(2023)\citenamefont {Burkov},
  \citenamefont {Smith}, \citenamefont {Hickey},\ and\ \citenamefont
  {Martin}}]{Burkov2023}%
  \BibitemOpen
  \bibfield  {author} {\bibinfo {author} {\bibfnamefont {A.~A.}\ \bibnamefont
  {Burkov}}, \bibinfo {author} {\bibfnamefont {M.}~\bibnamefont {Smith}},
  \bibinfo {author} {\bibfnamefont {A.}~\bibnamefont {Hickey}}, \ and\ \bibinfo
  {author} {\bibfnamefont {I.}~\bibnamefont {Martin}},\ }\href {\doibase
  10.1103/PhysRevB.108.205115} {\bibfield  {journal} {\bibinfo  {journal}
  {Phys. Rev. B}\ }\textbf {\bibinfo {volume} {108}},\ \bibinfo {pages}
  {205115} (\bibinfo {year} {2023})}\BibitemShut {NoStop}%
\bibitem [{\citenamefont {Xiao}\ \emph {et~al.}(2007)\citenamefont {Xiao},
  \citenamefont {Yao},\ and\ \citenamefont {Niu}}]{Xiao2007}%
  \BibitemOpen
  \bibfield  {author} {\bibinfo {author} {\bibfnamefont {D.}~\bibnamefont
  {Xiao}}, \bibinfo {author} {\bibfnamefont {W.}~\bibnamefont {Yao}}, \ and\
  \bibinfo {author} {\bibfnamefont {Q.}~\bibnamefont {Niu}},\ }\href {\doibase
  10.1103/PhysRevLett.99.236809} {\bibfield  {journal} {\bibinfo  {journal}
  {Phys. Rev. Lett.}\ }\textbf {\bibinfo {volume} {99}},\ \bibinfo {pages}
  {236809} (\bibinfo {year} {2007})}\BibitemShut {NoStop}%
\bibitem [{\citenamefont {Schaibley}\ \emph {et~al.}(2016)\citenamefont
  {Schaibley}, \citenamefont {Yu}, \citenamefont {Clark}, \citenamefont
  {Rivera}, \citenamefont {Ross}, \citenamefont {Seyler}, \citenamefont {Yao},\
  and\ \citenamefont {Xu}}]{Schaibley2016}%
  \BibitemOpen
  \bibfield  {author} {\bibinfo {author} {\bibfnamefont {J.~R.}\ \bibnamefont
  {Schaibley}}, \bibinfo {author} {\bibfnamefont {H.}~\bibnamefont {Yu}},
  \bibinfo {author} {\bibfnamefont {G.}~\bibnamefont {Clark}}, \bibinfo
  {author} {\bibfnamefont {P.}~\bibnamefont {Rivera}}, \bibinfo {author}
  {\bibfnamefont {J.~S.}\ \bibnamefont {Ross}}, \bibinfo {author}
  {\bibfnamefont {K.~L.}\ \bibnamefont {Seyler}}, \bibinfo {author}
  {\bibfnamefont {W.}~\bibnamefont {Yao}}, \ and\ \bibinfo {author}
  {\bibfnamefont {X.}~\bibnamefont {Xu}},\ }\href {\doibase
  10.1038/natrevmats.2016.55} {\bibfield  {journal} {\bibinfo  {journal}
  {Nature Reviews Materials}\ }\textbf {\bibinfo {volume} {1}},\ \bibinfo
  {pages} {16055} (\bibinfo {year} {2016})}\BibitemShut {NoStop}%
\bibitem [{\citenamefont {Vitale}\ \emph {et~al.}(2018)\citenamefont {Vitale},
  \citenamefont {Nezich}, \citenamefont {Varghese}, \citenamefont {Kim},
  \citenamefont {Gedik}, \citenamefont {Jarillo-Herrero}, \citenamefont
  {Xiao},\ and\ \citenamefont {Rothschild}}]{Vitale2018}%
  \BibitemOpen
  \bibfield  {author} {\bibinfo {author} {\bibfnamefont {S.~A.}\ \bibnamefont
  {Vitale}}, \bibinfo {author} {\bibfnamefont {D.}~\bibnamefont {Nezich}},
  \bibinfo {author} {\bibfnamefont {J.~O.}\ \bibnamefont {Varghese}}, \bibinfo
  {author} {\bibfnamefont {P.}~\bibnamefont {Kim}}, \bibinfo {author}
  {\bibfnamefont {N.}~\bibnamefont {Gedik}}, \bibinfo {author} {\bibfnamefont
  {P.}~\bibnamefont {Jarillo-Herrero}}, \bibinfo {author} {\bibfnamefont
  {D.}~\bibnamefont {Xiao}}, \ and\ \bibinfo {author} {\bibfnamefont
  {M.}~\bibnamefont {Rothschild}},\ }\href {\doibase
  https://doi.org/10.1002/smll.201801483} {\bibfield  {journal} {\bibinfo
  {journal} {Small}\ }\textbf {\bibinfo {volume} {14}},\ \bibinfo {pages}
  {1801483} (\bibinfo {year} {2018})}\BibitemShut {NoStop}%
\bibitem [{\citenamefont {Liu}\ \emph {et~al.}(2019)\citenamefont {Liu},
  \citenamefont {Gao}, \citenamefont {Zhang}, \citenamefont {He}, \citenamefont
  {Yu},\ and\ \citenamefont {Liu}}]{Liu2019}%
  \BibitemOpen
  \bibfield  {author} {\bibinfo {author} {\bibfnamefont {Y.}~\bibnamefont
  {Liu}}, \bibinfo {author} {\bibfnamefont {Y.}~\bibnamefont {Gao}}, \bibinfo
  {author} {\bibfnamefont {S.}~\bibnamefont {Zhang}}, \bibinfo {author}
  {\bibfnamefont {J.}~\bibnamefont {He}}, \bibinfo {author} {\bibfnamefont
  {J.}~\bibnamefont {Yu}}, \ and\ \bibinfo {author} {\bibfnamefont
  {Z.}~\bibnamefont {Liu}},\ }\href {\doibase 10.1007/s12274-019-2497-2}
  {\bibfield  {journal} {\bibinfo  {journal} {Nano Research}\ }\textbf
  {\bibinfo {volume} {12}},\ \bibinfo {pages} {2695} (\bibinfo {year}
  {2019})}\BibitemShut {NoStop}%
\bibitem [{\citenamefont {Yao}\ \emph {et~al.}(2008)\citenamefont {Yao},
  \citenamefont {Xiao},\ and\ \citenamefont {Niu}}]{Yao2008}%
  \BibitemOpen
  \bibfield  {author} {\bibinfo {author} {\bibfnamefont {W.}~\bibnamefont
  {Yao}}, \bibinfo {author} {\bibfnamefont {D.}~\bibnamefont {Xiao}}, \ and\
  \bibinfo {author} {\bibfnamefont {Q.}~\bibnamefont {Niu}},\ }\href {\doibase
  10.1103/PhysRevB.77.235406} {\bibfield  {journal} {\bibinfo  {journal} {Phys.
  Rev. B}\ }\textbf {\bibinfo {volume} {77}},\ \bibinfo {pages} {235406}
  (\bibinfo {year} {2008})}\BibitemShut {NoStop}%
\bibitem [{\citenamefont {Li}\ \emph {et~al.}(2020{\natexlab{b}})\citenamefont
  {Li}, \citenamefont {Wu}, \citenamefont {Feng}, \citenamefont {Guan},
  \citenamefont {Feng}, \citenamefont {Yao},\ and\ \citenamefont
  {Yang}}]{LiS2020}%
  \BibitemOpen
  \bibfield  {author} {\bibinfo {author} {\bibfnamefont {S.}~\bibnamefont
  {Li}}, \bibinfo {author} {\bibfnamefont {W.}~\bibnamefont {Wu}}, \bibinfo
  {author} {\bibfnamefont {X.}~\bibnamefont {Feng}}, \bibinfo {author}
  {\bibfnamefont {S.}~\bibnamefont {Guan}}, \bibinfo {author} {\bibfnamefont
  {W.}~\bibnamefont {Feng}}, \bibinfo {author} {\bibfnamefont {Y.}~\bibnamefont
  {Yao}}, \ and\ \bibinfo {author} {\bibfnamefont {S.~A.}\ \bibnamefont
  {Yang}},\ }\href {\doibase 10.1103/PhysRevB.102.235435} {\bibfield  {journal}
  {\bibinfo  {journal} {Phys. Rev. B}\ }\textbf {\bibinfo {volume} {102}},\
  \bibinfo {pages} {235435} (\bibinfo {year} {2020}{\natexlab{b}})}\BibitemShut
  {NoStop}%
\bibitem [{\citenamefont {Zhu}\ \emph {et~al.}(2023)\citenamefont {Zhu},
  \citenamefont {Zhang}, \citenamefont {Qin}, \citenamefont {Wang},
  \citenamefont {Wang}, \citenamefont {Miao}, \citenamefont {Liu},
  \citenamefont {Huang}, \citenamefont {Zhang}, \citenamefont {Xu},
  \citenamefont {Zhen}, \citenamefont {Chai},\ and\ \citenamefont
  {Xu}}]{Zhu2023}%
  \BibitemOpen
  \bibfield  {author} {\bibinfo {author} {\bibfnamefont {C.-Y.}\ \bibnamefont
  {Zhu}}, \bibinfo {author} {\bibfnamefont {Z.}~\bibnamefont {Zhang}}, \bibinfo
  {author} {\bibfnamefont {J.-K.}\ \bibnamefont {Qin}}, \bibinfo {author}
  {\bibfnamefont {Z.}~\bibnamefont {Wang}}, \bibinfo {author} {\bibfnamefont
  {C.}~\bibnamefont {Wang}}, \bibinfo {author} {\bibfnamefont {P.}~\bibnamefont
  {Miao}}, \bibinfo {author} {\bibfnamefont {Y.}~\bibnamefont {Liu}}, \bibinfo
  {author} {\bibfnamefont {P.-Y.}\ \bibnamefont {Huang}}, \bibinfo {author}
  {\bibfnamefont {Y.}~\bibnamefont {Zhang}}, \bibinfo {author} {\bibfnamefont
  {K.}~\bibnamefont {Xu}}, \bibinfo {author} {\bibfnamefont {L.}~\bibnamefont
  {Zhen}}, \bibinfo {author} {\bibfnamefont {Y.}~\bibnamefont {Chai}}, \ and\
  \bibinfo {author} {\bibfnamefont {C.-Y.}\ \bibnamefont {Xu}},\ }\href
  {\doibase 10.1038/s41467-023-38131-2} {\bibfield  {journal} {\bibinfo
  {journal} {Nature Communications}\ }\textbf {\bibinfo {volume} {14}},\
  \bibinfo {pages} {2521} (\bibinfo {year} {2023})}\BibitemShut {NoStop}%
\bibitem [{\citenamefont {Sangwan}\ \emph {et~al.}(2024)\citenamefont
  {Sangwan}, \citenamefont {Chica}, \citenamefont {Chu}, \citenamefont {Cheng},
  \citenamefont {Quintero}, \citenamefont {Hao}, \citenamefont {Mead},
  \citenamefont {Choi}, \citenamefont {Zu}, \citenamefont {Sheoran},
  \citenamefont {He}, \citenamefont {Liu}, \citenamefont {Qian}, \citenamefont
  {Laing}, \citenamefont {Kang}, \citenamefont {Gopalan}, \citenamefont
  {Wolverton}, \citenamefont {Dravid}, \citenamefont {Lauhon}, \citenamefont
  {Hersam},\ and\ \citenamefont {Kanatzidis}}]{Sangwan2024}%
  \BibitemOpen
  \bibfield  {author} {\bibinfo {author} {\bibfnamefont {V.~K.}\ \bibnamefont
  {Sangwan}}, \bibinfo {author} {\bibfnamefont {D.~G.}\ \bibnamefont {Chica}},
  \bibinfo {author} {\bibfnamefont {T.-C.}\ \bibnamefont {Chu}}, \bibinfo
  {author} {\bibfnamefont {M.}~\bibnamefont {Cheng}}, \bibinfo {author}
  {\bibfnamefont {M.~A.}\ \bibnamefont {Quintero}}, \bibinfo {author}
  {\bibfnamefont {S.}~\bibnamefont {Hao}}, \bibinfo {author} {\bibfnamefont
  {C.~E.}\ \bibnamefont {Mead}}, \bibinfo {author} {\bibfnamefont
  {H.}~\bibnamefont {Choi}}, \bibinfo {author} {\bibfnamefont {R.}~\bibnamefont
  {Zu}}, \bibinfo {author} {\bibfnamefont {J.}~\bibnamefont {Sheoran}},
  \bibinfo {author} {\bibfnamefont {J.}~\bibnamefont {He}}, \bibinfo {author}
  {\bibfnamefont {Y.}~\bibnamefont {Liu}}, \bibinfo {author} {\bibfnamefont
  {E.}~\bibnamefont {Qian}}, \bibinfo {author} {\bibfnamefont {C.~C.}\
  \bibnamefont {Laing}}, \bibinfo {author} {\bibfnamefont {M.-A.}\ \bibnamefont
  {Kang}}, \bibinfo {author} {\bibfnamefont {V.}~\bibnamefont {Gopalan}},
  \bibinfo {author} {\bibfnamefont {C.}~\bibnamefont {Wolverton}}, \bibinfo
  {author} {\bibfnamefont {V.~P.}\ \bibnamefont {Dravid}}, \bibinfo {author}
  {\bibfnamefont {L.~J.}\ \bibnamefont {Lauhon}}, \bibinfo {author}
  {\bibfnamefont {M.~C.}\ \bibnamefont {Hersam}}, \ and\ \bibinfo {author}
  {\bibfnamefont {M.~G.}\ \bibnamefont {Kanatzidis}},\ }\href {\doibase
  10.1126/sciadv.ado8272} {\bibfield  {journal} {\bibinfo  {journal} {Science
  Advances}\ }\textbf {\bibinfo {volume} {10}},\ \bibinfo {pages} {eado8272}
  (\bibinfo {year} {2024})}\BibitemShut {NoStop}%
\bibitem [{\citenamefont {Zhu}\ \emph {et~al.}()\citenamefont {Zhu},
  \citenamefont {Zhu}, \citenamefont {Qin}, \citenamefont {He}, \citenamefont
  {Yue}, \citenamefont {Huang}, \citenamefont {Li}, \citenamefont {Sun},
  \citenamefont {Ye}, \citenamefont {Du}, \citenamefont {Sui}, \citenamefont
  {Li}, \citenamefont {Mao}, \citenamefont {Zhen},\ and\ \citenamefont
  {Xu}}]{Zhu2024}%
  \BibitemOpen
  \bibfield  {author} {\bibinfo {author} {\bibfnamefont {B.-X.}\ \bibnamefont
  {Zhu}}, \bibinfo {author} {\bibfnamefont {C.-Y.}\ \bibnamefont {Zhu}},
  \bibinfo {author} {\bibfnamefont {J.-K.}\ \bibnamefont {Qin}}, \bibinfo
  {author} {\bibfnamefont {W.}~\bibnamefont {He}}, \bibinfo {author}
  {\bibfnamefont {L.-Q.}\ \bibnamefont {Yue}}, \bibinfo {author} {\bibfnamefont
  {P.-Y.}\ \bibnamefont {Huang}}, \bibinfo {author} {\bibfnamefont
  {D.}~\bibnamefont {Li}}, \bibinfo {author} {\bibfnamefont {R.-Y.}\
  \bibnamefont {Sun}}, \bibinfo {author} {\bibfnamefont {S.}~\bibnamefont
  {Ye}}, \bibinfo {author} {\bibfnamefont {Y.}~\bibnamefont {Du}}, \bibinfo
  {author} {\bibfnamefont {J.-H.}\ \bibnamefont {Sui}}, \bibinfo {author}
  {\bibfnamefont {M.-Y.}\ \bibnamefont {Li}}, \bibinfo {author} {\bibfnamefont
  {J.}~\bibnamefont {Mao}}, \bibinfo {author} {\bibfnamefont {L.}~\bibnamefont
  {Zhen}}, \ and\ \bibinfo {author} {\bibfnamefont {C.-Y.}\ \bibnamefont
  {Xu}},\ }\href {\doibase https://doi.org/10.1002/inf2.12600} {\bibfield
  {journal} {\bibinfo  {journal} {InfoMat}\ }\textbf {\bibinfo {volume}
  {n/a}},\ \bibinfo {pages} {e12600}}\BibitemShut {NoStop}%
\bibitem [{SMs()}]{SMsnp2se6}%
  \BibitemOpen
  \href@noop {} {\ }\bibinfo {note} {See Supplemental Material for calculation
  details, electronic properties of the SnP$_{2}$Se$_{6}$ monolayer using the
  HSE functional, atomic and electronic properties of SnP$_{2}$Se$_{6}$
  few-layers, electronic properties of the GeP$_{2}$S$_{6}$
  monolayer.}\BibitemShut {Stop}%
\bibitem [{\citenamefont {Lin}\ \emph {et~al.}(2019)\citenamefont {Lin},
  \citenamefont {Liu}, \citenamefont {Wu}, \citenamefont {Cheng}, \citenamefont
  {Liu}, \citenamefont {Cho}, \citenamefont {Wang},\ and\ \citenamefont
  {Lu}}]{LinM2019}%
  \BibitemOpen
  \bibfield  {author} {\bibinfo {author} {\bibfnamefont {M.}~\bibnamefont
  {Lin}}, \bibinfo {author} {\bibfnamefont {P.}~\bibnamefont {Liu}}, \bibinfo
  {author} {\bibfnamefont {M.}~\bibnamefont {Wu}}, \bibinfo {author}
  {\bibfnamefont {Y.}~\bibnamefont {Cheng}}, \bibinfo {author} {\bibfnamefont
  {H.}~\bibnamefont {Liu}}, \bibinfo {author} {\bibfnamefont {K.}~\bibnamefont
  {Cho}}, \bibinfo {author} {\bibfnamefont {W.-H.}\ \bibnamefont {Wang}}, \
  and\ \bibinfo {author} {\bibfnamefont {F.}~\bibnamefont {Lu}},\ }\href
  {\doibase https://doi.org/10.1016/j.apsusc.2019.07.115} {\bibfield  {journal}
  {\bibinfo  {journal} {Applied Surface Science}\ }\textbf {\bibinfo {volume}
  {493}},\ \bibinfo {pages} {1334} (\bibinfo {year} {2019})}\BibitemShut
  {NoStop}%
\bibitem [{\citenamefont {Perdew}\ \emph {et~al.}(1996)\citenamefont {Perdew},
  \citenamefont {Burke},\ and\ \citenamefont {Ernzerhof}}]{Perdew1996}%
  \BibitemOpen
  \bibfield  {author} {\bibinfo {author} {\bibfnamefont {J.~P.}\ \bibnamefont
  {Perdew}}, \bibinfo {author} {\bibfnamefont {K.}~\bibnamefont {Burke}}, \
  and\ \bibinfo {author} {\bibfnamefont {M.}~\bibnamefont {Ernzerhof}},\ }\href
  {\doibase 10.1103/PhysRevLett.77.3865} {\bibfield  {journal} {\bibinfo
  {journal} {Phys. Rev. Lett.}\ }\textbf {\bibinfo {volume} {77}},\ \bibinfo
  {pages} {3865} (\bibinfo {year} {1996})}\BibitemShut {NoStop}%
\bibitem [{\citenamefont {Mera~Acosta}\ \emph {et~al.}(2021)\citenamefont
  {Mera~Acosta}, \citenamefont {Yuan}, \citenamefont {Dalpian},\ and\
  \citenamefont {Zunger}}]{Acosta2021}%
  \BibitemOpen
  \bibfield  {author} {\bibinfo {author} {\bibfnamefont {C.}~\bibnamefont
  {Mera~Acosta}}, \bibinfo {author} {\bibfnamefont {L.}~\bibnamefont {Yuan}},
  \bibinfo {author} {\bibfnamefont {G.~M.}\ \bibnamefont {Dalpian}}, \ and\
  \bibinfo {author} {\bibfnamefont {A.}~\bibnamefont {Zunger}},\ }\href
  {\doibase 10.1103/PhysRevB.104.104408} {\bibfield  {journal} {\bibinfo
  {journal} {Phys. Rev. B}\ }\textbf {\bibinfo {volume} {104}},\ \bibinfo
  {pages} {104408} (\bibinfo {year} {2021})}\BibitemShut {NoStop}%
\bibitem [{\citenamefont {Xie}\ and\ \citenamefont {Cui}(2016)}]{Lu2016}%
  \BibitemOpen
  \bibfield  {author} {\bibinfo {author} {\bibfnamefont {L.}~\bibnamefont
  {Xie}}\ and\ \bibinfo {author} {\bibfnamefont {X.}~\bibnamefont {Cui}},\
  }\href {\doibase 10.1073/pnas.1523012113} {\bibfield  {journal} {\bibinfo
  {journal} {Proceedings of the National Academy of Sciences}\ }\textbf
  {\bibinfo {volume} {113}},\ \bibinfo {pages} {3746} (\bibinfo {year}
  {2016})}\BibitemShut {NoStop}%
\bibitem [{\citenamefont {Heyd}\ \emph {et~al.}(2003)\citenamefont {Heyd},
  \citenamefont {Scuseria},\ and\ \citenamefont {Ernzerhof}}]{Heyd2003}%
  \BibitemOpen
  \bibfield  {author} {\bibinfo {author} {\bibfnamefont {J.}~\bibnamefont
  {Heyd}}, \bibinfo {author} {\bibfnamefont {G.~E.}\ \bibnamefont {Scuseria}},
  \ and\ \bibinfo {author} {\bibfnamefont {M.}~\bibnamefont {Ernzerhof}},\
  }\href {\doibase 10.1063/1.1564060} {\bibfield  {journal} {\bibinfo
  {journal} {The Journal of Chemical Physics}\ }\textbf {\bibinfo {volume}
  {118}},\ \bibinfo {pages} {8207} (\bibinfo {year} {2003})}\BibitemShut
  {NoStop}%
\end{thebibliography}
\end{document}